\documentclass[twocolumn]{aastex6}
\usepackage{xcolor}

\def\jnlref#1{{\rm#1}}
\def\apj{\jnlref{ApJ}}
\def\apjs{\jnlref{ApJS}}
\def\aap{\jnlref{A\&A}}
\def\mnras{\jnlref{MNRAS}}

\begin{document}

%% LaTeX will automatically break titles if they run longer than
%% one line. However, you may use \\ to force a line break if
%% you desire.

\title{On the evolution of galaxy spin in a cosmological hydrodynamic simulation of galaxy clusters}
\shorttitle{on the evolution of galaxy spin}

\author{Hoseung Choi\altaffilmark{1} and Sukyoung K. Yi\altaffilmark{1}}
\affil{$^1$Department of Astronomy and Yonsei University Observatory, Yonsei University, 50 Yonsei-ro, Seodaemun-gu, Seoul 03722, Republic of Korea: choi.h@yonsei.ac.kr, yi@yonsei.ac.kr}

%% Notice that each of these authors has alternate affiliations, which
%% are identified by the \altaffilmark after each name.  Specify alternate
%% affiliation information with \altaffiltext, with one command per each
%% affiliation.

%% Mark off the abstract in the ``abstract'' environment. 
\begin{abstract}
 The traditional view of the morphology-spin connection is being challenged by recent integral-field-unit observations, as the majority of early-type galaxies are found to have a rotational component that is often as large as a dispersion component. 
Mergers are often suspected to be critical in galaxy spin evolution, yet the details of their roles are still unclear.
 We present the first results on the spin evolution of galaxies in cluster environments through a cosmological hydrodynamic simulation. 
 Galaxies spin down globally with cosmic evolution.
 Major (mass ratios $>$ 1/4) and minor (1/4 $\geq$  mass ratios $>$ 1/50) mergers are important 
contributors to the spin down in particular in massive galaxies. 
 Minor mergers appear to have stronger cumulative effects than major mergers. 
 Surprisingly, the dominant driver of galaxy spin down seems to be environmental effects rather than mergers. However, since
 multiple processes act in combination, it is difficult to separate their individual roles.
 We briefly discuss the caveats and future studies that are called for.
\end{abstract}

%% Keywords should appear after the \end{abstract} command. 
%% See the online documentation for the full list of available subject
%% keywords and the rules for their use.

%% From the front matter, we move on to the body of the paper.
%% Sections are demarcated by \section and \subsection, respectively.
%% Observe the use of the LaTeX \label
%% command after the \subsection to give a symbolic KEY to the
%% subsection for cross-referencing in a \ref command.
%% You can use LaTeX's \ref and \label commands to keep track of
%% cross-references to sections, equations, tables, and figures.
%% That way, if you change the order of any elements, LaTeX will
%% automatically renumber them.

%% We recommend that authors also use the natbib \citep
%% and \citet commands to identify citations.  The citations are
%% tied to the reference list via symbolic KEYs. The KEY corresponds
%% to the KEY in the \bibitem in the reference list below. 

\section{Introduction} \label{sec:intro}
Ever since the first classification of ``galaxies'' by \citet{Hubble1926}, the origin of galaxy morphology has been a main quest in 
astrophysics. 
The prevailing interpretation has been in terms of kinematic properties: elliptical galaxies are dispersion-dominated
\citep{Bertola1975,Binney1976,Illingworth1977}, while disk galaxies are rotation dominated. 
This was viewed as reasonable in the current-favorite hierarchical merger paradigm, if elliptical galaxies are merger remnants where the 
angular momentum of merging disks is likely reduced after repeated mergers. 

Galaxy investigations have often been based on galaxy morphology, but this may have to change. 
Recent integral field unit (IFU) spectroscopic observations have revealed that elliptical galaxies, too, 
contain a substantial fraction of rotating component, in contrast to the textbook expectation. 
More than three quarters of the ATLAS$^{\rm 3D}$ early-type galaxies are classified as fast rotators \citep{Emsellem2011}, damaging the traditional understanding of the morphology-spin connection.
Spin\footnote{See Section 2.4 for its definition.} is suggested 
to be a representative property of a galaxy, transforming the way galaxy studies are performed. 
It is essential to develop an understanding on the spin evolution of galaxies to face this paradigm shift.

Galaxies probably form first as a rotating disk because the collision among pre-galactic clouds would not likely achieve a zero net angular 
momentum, as suggested by tidal torque theory \citep{Peebles1969}. 
Numerical simulations demonstrated that galaxy mergers may result in dispersion-dominated morphologically-elliptical galaxies under centain conditions
\citep[e.g.][]{Toomre1977, Gerhard1981, Barnes1988, Barnes1992, Hernquist1992, Naab2006, Cox2006}.
On the other hand, recent studies found that the final morphology of a galaxy depends more on its individual gas accretion history rather 
than galaxy merger history \citep{Sales2012}.
Morphology and spin seem linked \citep{Fall1983,Fall2013}, but the role of mergers and the details of the spin evolution are far from 
being clear.

Theoretical attempts have been made to explain the origin of slow and fast rotators. 
A semi-analytic calculation succeeded first in reproducing the ratio between slow and fast rotators in a cosmological context \citep{Khochfar2011}. 
Hydrodynamic simulations were also employed to challenge this problem. 
Idealized equal-mass merger simulations were used to demonstrate that merger remnants can approximate the spin-ellipticity distribution of galaxies observed \citep{Bois2011}. 
A cosmological zoom-in hydrodynamic simulation on 44 central galaxies reproduced massive round slow rotators through major and minor mergers (\citealp{Naab2014}; see also \citealp{Moody2014}).
These studies made important advances, and we are now ready to explore the issue in greater detail covering a much wider parameter space in merger conditions. 
In this paper we examine the evolutions of thousands of galaxies in simulated clusters. 
We aim at understanding the general trend of galaxy spin evolution with a much larger sample and a wider range of merger 
histories compared to the earlier studies, while maintaining the simulation resolution reasonably high.

\section{Methods}
\subsection{Numerical Simulations} \label{sec:simulations}
We used the adaptive mesh refinement code RAMSES \citep{Teyssier2002} to perform a cosmological zoom-in simulation on galaxy cluster 
scales. 
We first ran a dark matter-only cosmological simulation for a cube 200\,h$^{-1}\,Mpc$ on a side to identify target clusters. 
We selected 16 dense regions of varying virial mass $\rm{M}_{200} = 10^{13.5} - 10^{15} \rm{M}_{\odot}$ from the cube and conducted a zoom-in simulation on these regions (out to 3\,$\rm{R}_{200}$), this time including hydrodynamic calculations as well. 
An up-to-date baryon physics recipe including AGN feedback and supernova feedback \citep{Dubois2012} was adopted.
Throughout the simulation and analyses, we assumed the WMAP7 cosmology \citep{Komatsu2011}: $\rm \Omega_{m} = 0.272, \Omega_{\Lambda}=0.728, H_{0} = 70.4km\,s^{-1}\,Mpc^{-1}, \sigma_{8}=0.809$, and $\rm n = 0.963$.

We allowed the smallest cell to be 380\,$\rm h^{-1}\,pc$ in size, but only the centers of a small number of galaxies actually reached to the maximum refinement level. 
Note that the Horizon-AGN simulation \citep{Dubois2014a} has the same DM particle mass resolution but allowed up to 760\,$\rm h^{-1}\,pc$ cell resolution refinement. 
In practice, the effective simulation resolution is 760\,$\rm h^{-1}\,pc$.
Still, the resolution is comparable to those of other state-of-the-art simulations \citep{Vogelsberger2014a,Schaye2015,Khandai2015}. 
One minor difference that emerges from allowing a small number of higher resolution cells with the same DM particle mass resolution is that the stellar mass in the simulation is $1/8$ of that of the Horizon-AGN simulation because the stellar mass depends on the maximum level of refinement allowed.

\subsection{Galaxy Identification} \label{sec:galaxydetection}
Galaxies were identified by AdaptaHOP halo finder \citep{Aubert2004} 
using the most massive sub-node method \citep{Tweed2009} for substructures. 
Following \citet{Dubois2014a}, 178 times the mean total matter density is used as the local density threshold for galaxy detection.
More than 200 stellar particles are needed for a galaxy to be identified, roughly corresponding to $\rm{M}_{*} = 10^{8}M_{\odot}$. 
Minimum size of a galaxy is 1\,$\rm h^{-1}\,kpc$, larger than the minimum grid size to avoid the effect of numerical noise.

While the 200 stellar particle criterion is sufficient to robustly identify galaxies, we selected only galaxies with $M_{*} > 5 \times 10^{9}\rm M_{\odot}$ at redshift $z = 0$ for the main sample.
At this mass cut, the minimum number of stellar particles is $\approx 10^4$, enabling reliable measurement of galaxy rotation in radial 
bins.
The main sample with the mass cut consists of 1726 galaxies. 
The other, smaller galaxies, were only considered as merging satellites.
Because each galaxy has an explicit particle membership, stellar particles that belong to merging satellites were ignored when we calculated the rotational property of the main galaxy.
Among 1726 galaxies, 211 are hosted by main halos, and 1515 galaxies are hosted by satellite halos as defined by AdaptaHOP algorithm. 
And there are 891 galaxies inside 1\,$\rm{R}_{200}$ of clusters.

\subsection{Merger Trees and Merger Definition} \label{sec:Merger}

The galaxy progenitor-descendant relation is determined by monitoring particle transfer between two consecutive snapshots. 
After determining the progenitor-descendant relation, we use ConsistentTrees \citep{Behroozi2013a} to correct minor errors such as trees being disconnected when a galaxy passes through a larger galaxy. 
However, the ability of ConsistentTrees to predict the gravitational evolution of DM halos is not used in this analysis because we are mainly interested in the merger trees of galaxies, not halos.
We found that the resulting galaxy merger trees are insensitive to the choice of tree building parameters mainly because our galaxies are defined by a sufficiently large number of particles.

It is worth mentioning here that ``merger'' in this study has a narrower meaning than the conventional halo merger. We define the beginning 
of a merger as when two galaxies start interacting. For comparison, halo mergers are conventionally assumed to start when the secondary halo 
crosses the virial radius of the primary halo. In practice, the beginning of a merger in our analysis happens when the 
distance between two galaxies becomes smaller than 10 times the sum of the effective radii of the two galaxies and does not 
grow larger. The end of the merger is defined by the final coalescence of the two galaxies, or the time when the 
secondary galaxy is no longer detected by the halo finder.

The merger mass ratio is defined as the stellar mass ratio at the beginning of the merger. 
In this study we classify mergers with mass ratio ($M_{\rm secondary}/M_{\rm primary} \geq 1/4$) as major mergers, 
and ($1/4 > M_{\rm secondary}/M_{\rm primary} \geq 1/50$) as minor mergers. 
Note that compared to the usual minor merger limit of 1/10, we include even smaller mergers as minor mergers.
We expect this will help us to estimate the upper limit of the merger effects.
The lower limit of mass ratio is 1/50, as the smallest satellite galaxy we can identify is 1/50 of the smallest main sample galaxy.

\subsection{Galaxy properties} \label{sec:galProp}

Galaxies identified by a halo finder sometimes have peculiar shapes. 
This is especially true for the brightest cluster galaxies (BCG) with a large part of the intra cluster light (ICL) components attached to them.
There is no clear cut in distinguishing the BCG component from ICL components. 
To reduce possible contamination from background stellar components, we adopt a stellar surface density cut of $\rm \Sigma_{\ast} > 10^{6} M_{\odot} kpc^{-2}$. 
Then the effective radius R$_{eff}$ is defined as the radius containing half of the stellar mass above the surface density cut.

It is important to precisely determine the center of mass and the center of velocity to calculate kinematic properties.
The galaxy center is defined as the three-dimensional density peak of stellar particles. 
The system velocity of a galaxy is defined as the average velocity of the stellar particles inside one effective radius. 
Since a host galaxy does not include stellar particles of satellite galaxies, this leads to a robust estimation of the system velocity. 

The rotation parameter is the central quantity in this study. 
It is defined following \citet{Emsellem2007} as:
\begin{equation}\label{eq:1}
\lambda_R \equiv \frac{\Sigma_{i=1}^{N_p} F_i R_i
  |V_i|}{\Sigma_{i=1}^{N_p}F_i R_i \sqrt{V_i^2 +\sigma_i^2}}, 
\end{equation}
where $\rm{V}_{i}$ is the velocity along the projection axis, and $\sigma_{i}$
is the velocity dispersion at a given pixel.
In practice, $\lambda_R$ and galaxy ellipticity $\epsilon$ are measured based on the stellar density map and velocity map projected along 
the original orientation of a galaxy given in the simulation to account for the random orientation of observed galaxies.
We have tried giving random projection effects as well, but it did not make any noticeable difference to our results.

The summation in the equation \ref{eq:1} runs over each bin of concentric ellipses with a common $\epsilon$ and a position angle(PA).
The PA and $\epsilon$ of a galaxy are measured using the publicly available MGE package \citep{Emsellem1994,Cappellari2002}. 
Although $\epsilon$ and PA can vary with radius, we fix $\epsilon$ and PA as measured at $\sqrt{ab} = R_{eff}$, where $a$ and $b$ are the 
semi major axis and semi minor axis. It should be noted that we mean the mass-weighted rotational property of a galaxy inside 1 effective 
radius by ``spin'' throughout this paper. This is what is practically measurable by current observations.

\section{Results} \label{sec:Results}
\begin{figure}[t!]
\figurenum{1}
\includegraphics[width=0.47\textwidth]{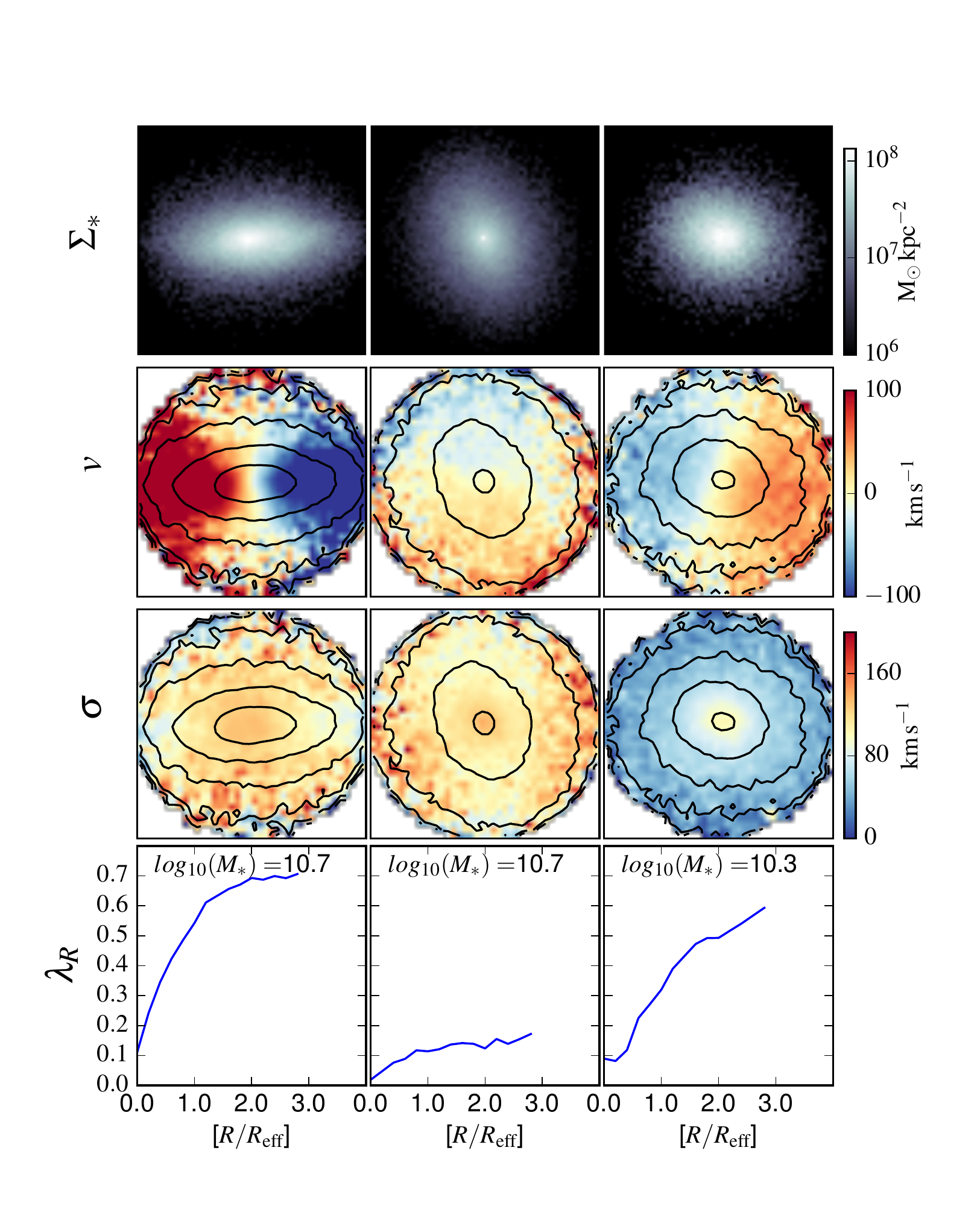}
\caption{Three examples of model galaxies at $z=0$. 
First row: projected stellar density within 4\,R$_{eff}$ in log scale.
Second row: mass weighted radial velocity map of stellar component with projected stellar density contour. 
Third row: mass weighted radial velocity dispersion of stellar component with projected stellar density contour.
Last row: radial profile of rotation parameter.
Most of the model galaxies in our clusters are early-type in morphology, yet their kinematic properties vary significantly, as found in recent IFU spectroscopic observations.
\label{fig:f1}}
\end{figure}

Most of the galaxies in our clusters appear to be early-type in morphology, as expected from the morphology-density relation \citep{Dressler1980}. 
Figure \ref{fig:f1} presents three examples of our model galaxies. 
They all appear to be early-type in projected stellar density (top), but their kinematic properties are markedly different. 
The ordered rotational velocity (2nd row) is high in two galaxies (A and C), while velocity dispersion (3rd row) is unusually low in C for an elliptical galaxy. 
We present the rotation parameter as a function of radial distance from the galaxy center (bottom) defined as equation \ref{eq:1}. 
Galaxies A and C have a fast rotator profile, whereas B shows a typical profile of a slow rotator.

\begin{figure}[t!]
\figurenum{2}
\includegraphics[width=0.47\textwidth]{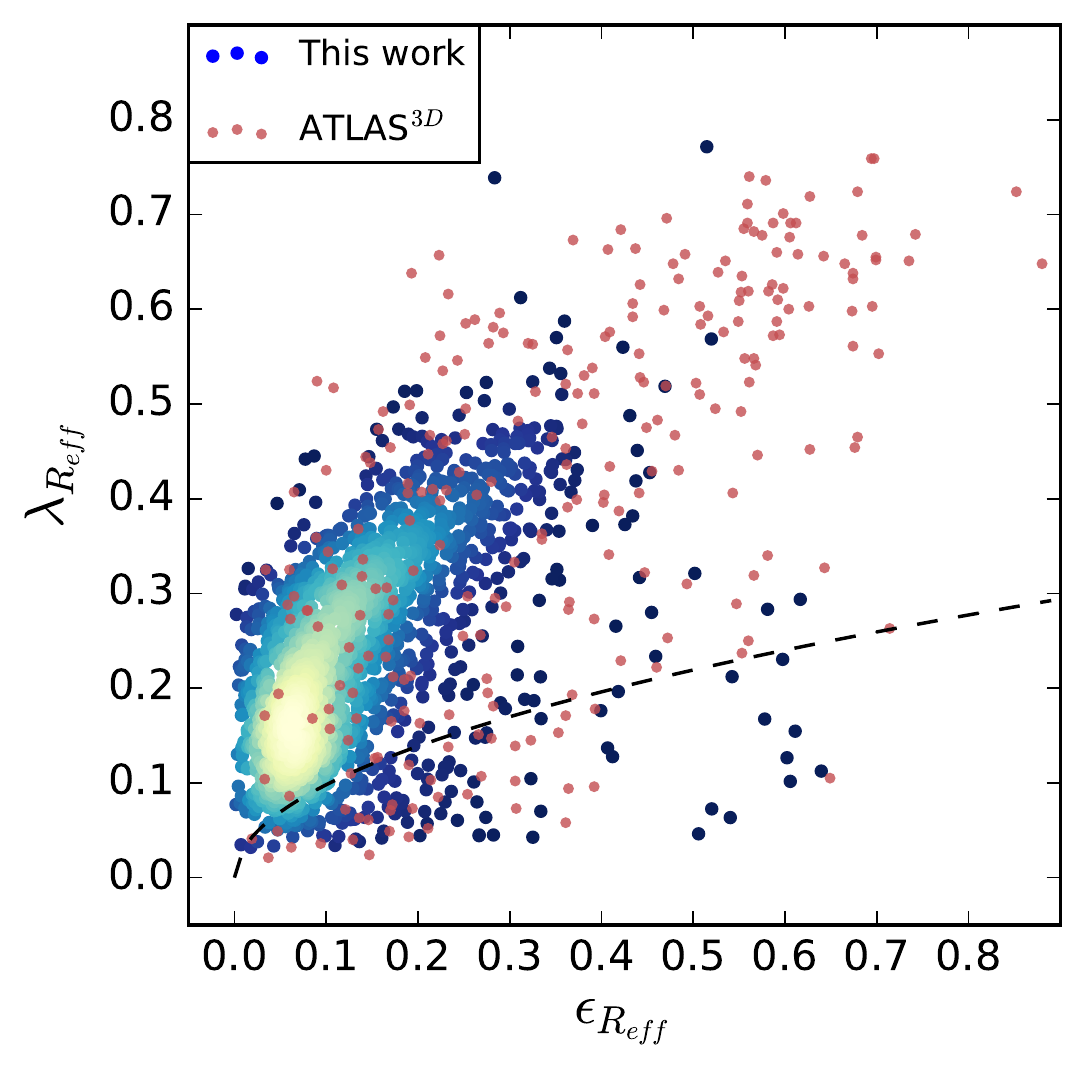}
\caption{Rotation parameter vs. ellipticity. 
The model galaxies in our clusters at $z=0$ (blue dots and contours) are presented in comparison to the ATLAS$^{\rm 3D}$ data (red dots). 
All measurements were performed at 1R$_{eff}$. 
The black dashed line shows the empirical demarcation between slow (below) and fast (above) rotators \citep{Emsellem2011}.
\label{fig:f2}}
\end{figure}

Figure \ref{fig:f2} shows the model galaxies in comparison with the ATLAS$^{\rm 3D}$ observational data in the rotation parameter vs. ellipticity plane. 
In a simplistic case of circular symmetry, $\lambda_{\rm{Reff}} \approx$  0.1 if $\sigma =10\,v_{\rm{rot}}$ and $\lambda_{\rm{Reff}} \approx 0.7$ if $\sigma = v_{\rm{rot}}$,
while $\lambda_{\rm{Reff}} = 0.31 \sqrt{\epsilon}$ (dashed curve) has been suggested as a demarcation line between slow and fast rotators. 
The observational data and models consistently suggest a positive correlation between ellipticity and spin, in the sense that rounder galaxies tend to rotate more slowly. 
They present a similar sequence in the diagram, but the ATLAS$^{\rm 3D}$ data span a larger area in the parameter space. 
Our models are rounder and more slowly rotating than the ATLAS$^{\rm 3D}$ sample. 
This may be partly due to the fact that our galaxies are exclusively in clusters where environmental effects are more dramatic. Of the 1726 model galaxies, 
138 (8\%) are classified as slow rotators, which is in rough agreement with the observations. 
The early SAURON data appeared to suggest a separated distribution between slow and fast rotators \citep{Emsellem2007}, but our data do not present such a dichotomy.

\begin{figure}[t!]
\figurenum{3}
\includegraphics[width=0.47\textwidth]{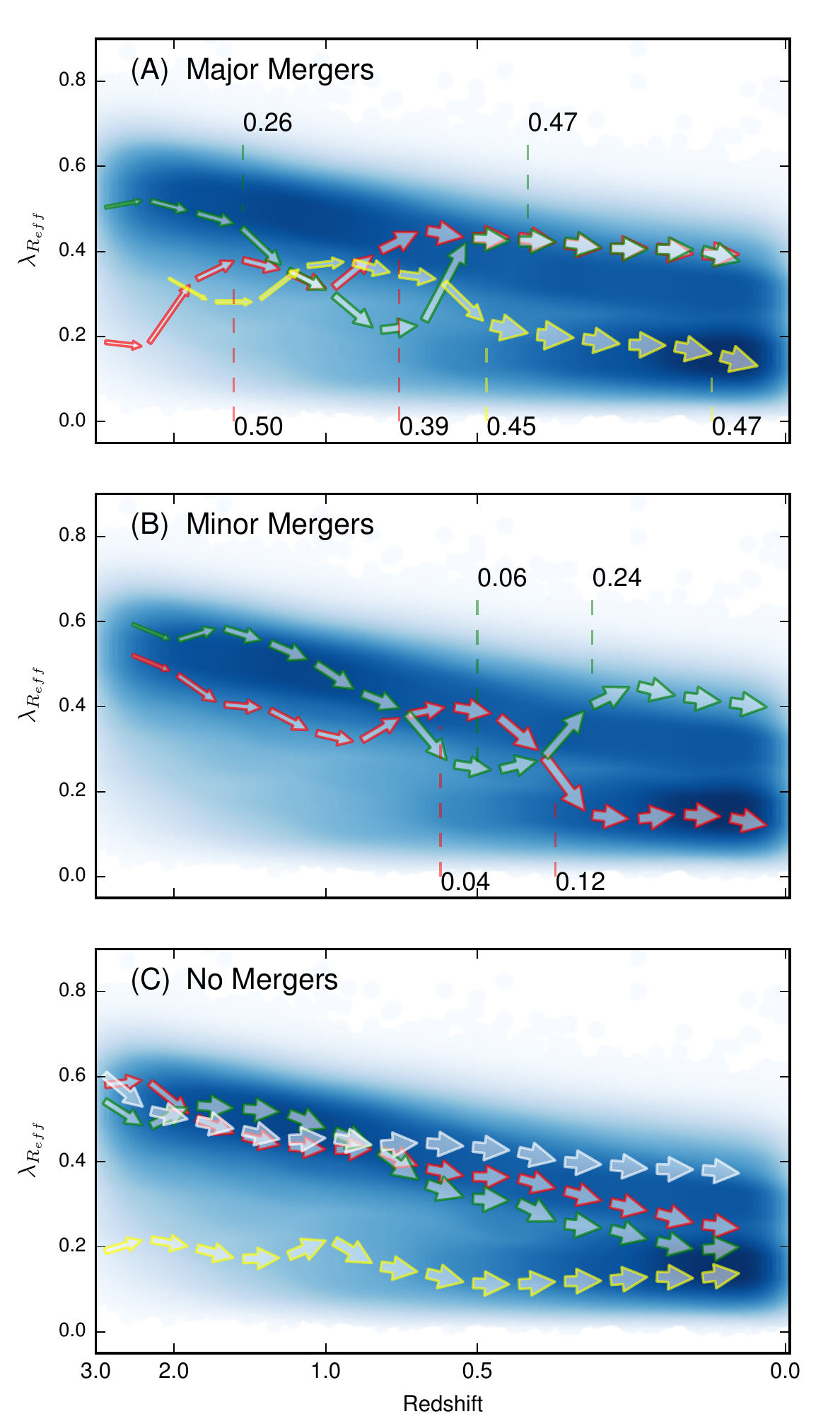}
\caption{Rotation parameter density (blue shades) of model cluster galaxies since $z=3$. 
Top panel shows three individual galaxies (arrows) that experienced major mergers during this period of time. 
The moments of mergers are marked with the mass ratio. 
The size of the arrow follows the stellar mass of each galaxy and is normalized at $z=0$. 
Middle panel shows the galaxies that only had minor mergers, and bottom panel shows the galaxies that did not experience any merger since $z=3$.
In each panel an identical density shade is given for comparison purposes.
\label{fig:f3}}
\end{figure}

We present the evolution of  galaxy rotation parameter starting from $z=3$, 
because it is non-trivial to extract the galaxy merger history accurately from a cosmological simulation at higher redshifts where galaxy interactions and mergers occurred in more chaotic manner. 
At $z=3$, most of the model galaxies were fast rotators (Figure \ref{fig:f3}). 
It seems that galaxies are indeed born generally as rotating disks as tidal torque theory suggests. 
Galaxy mass growth was quick in early high-density peaks which later evolve to be galaxy clusters, 
and thus the cluster galaxies reached a stage where dispersion is more important than rotation ($\lambda < 0.7$) already at $z=3$. 
The peak in the rotation parameter distribution moves from roughly 0.6 at $z=3$ to 0.2 at $z=0$. 
The decline is almost steady and global. 
There seem to be two (upper and lower) sequences, but actual changes of galaxy 
rotation parameter are complicated. 
The transition from fast to slow rotators occurs gradually.

Figure \ref{fig:f3}(A) shows the evolutionary paths of three sample galaxies that had major ($M_{\rm secondary}/M_{\rm primary} \geq 1/4$) mergers since $z=3$. 
When a galaxy experiences a major merger, its impact on rotation is not simple to predict. 
Depending on the merger conditions and the properties of merging galaxies (e.g. orbital parameters, gas fraction, and 
environment), 
the merger can enhance or reduce spin with different magnitudes. The same is true for minor mergers. 
Panel B shows two sample galaxies that had minor mergers since $z=3$; their impact on spin is similarly complex. 
Note that we include all the mergers with $1/50 \leq M_{\rm secondary}/M_{\rm primary} < 1/4$ in the ``minor'' merger category 
so that we consider the effect of virtually all minor mergers even beyond what is usually possible for observational detection. 
Panel C shows the galaxies that apparently did not have any merger event since $z=3$. 
To our surprise and against general expectation, the spin decline is possible and even clearer without mergers. 

\begin{figure}[t!]
\figurenum{4}
\includegraphics[width=0.47\textwidth]{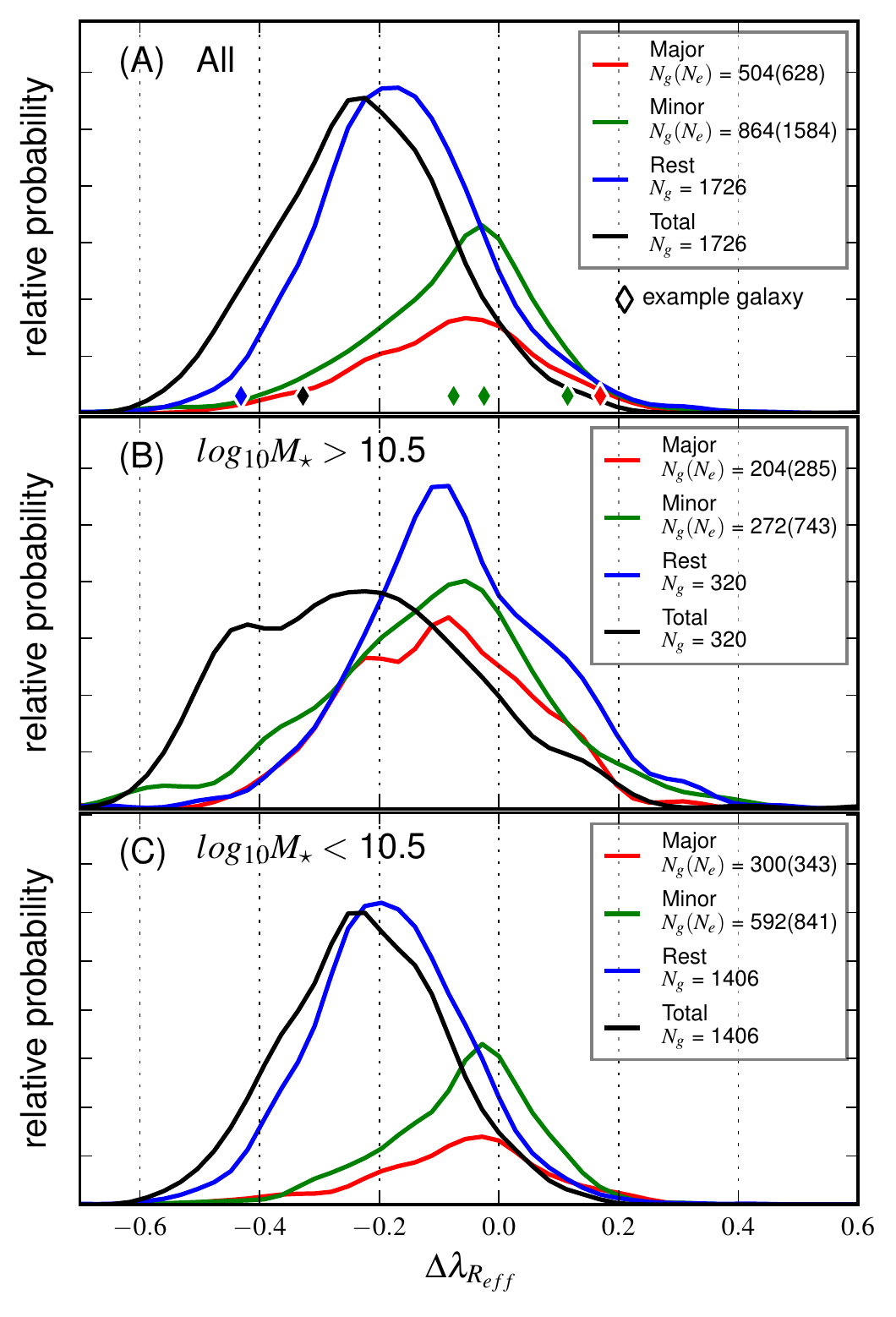}
\caption{Relative contribution of major mergers (red line), minor mergers (green), 
``Rest'' (blue) and the total (black) spin changes of the 1726 galaxies at z=0. 
(A) For all galaxies. 
The red curve shows the distribution of spin change caused by 628 major merger events which occurred in 504 galaxies. 
The three curves cannot simply be summed to match the total (black) because the distributions for the mergers are for individual merger events and a galaxy can have multiple mergers. 
The case of a sample galaxy which had one major merger and 3 minor mergers is shown by diamonds. 
(B) The case for massive galaxies of log $\rm{M}_{\ast}/{M}_{\odot} >$ 10.5. 
(C) The case for less massive galaxies of 9.7 $\leq$ log($\rm{M}_{\ast}/{M}_{\odot})$ $<$ 10.5 
shown by diamonds. 
\label{fig:f4}}
\end{figure}

The significance of mergers and other processes on spin can be quantified. 
We monitored the galaxy merger history and measured the spin change $\Delta\lambda$ between before and after each merger. 
Here, $\lambda$ before the merger is the average over the 0.5\,Gyr window just prior to the 
beginning of the merger as defined in Section \ref{sec:Merger}. On the other hand, $\lambda$ after the merger is the average over 0.5\,Gyr 
starting from 0.5\,Gyr after the final coalescence to let the merger remnant settle down.
The positions and widths of the windows were empirically chosen, and our results do not change much with the choice. 
Figure \ref{fig:f4} shows the relative contribution of major mergers, minor mergers, and ``Rest'', where
\begin{equation}
\Delta\lambda_{total} = \sum_{i} \Delta\lambda_{Major,i} + \sum_{j}\Delta\lambda_{minor,j} + \Delta\lambda_{Rest}.
\end{equation}
The total (combined) change of spin since $z=3$ in each galaxy is typically negative (spin down) with $\Delta\lambda \approx -0.25$. 

In our sample of 1726 galaxies, 504 galaxies experienced 628 major mergers, hence on average only 1.2 major mergers per galaxy. 
Major mergers have a wide variety of (both positive and negative) impact on spin 
depending on the details of the merger, but on average reduces galaxy spin. 
The peak of the distribution is only marginally on the negative side and the amplitude is low, 
which means that major mergers are not to be ignored, but not the dominant driver of the galaxy spin evolution. 

Minor mergers have a larger impact mainly because they are more frequent, but their impact (position and amplitude of the peak) is still 
modest. 
Compared to major mergers, minor mergers involve companions of smaller masses but are still effective in reducing galaxy spin 
over a long period of time because only a small parameter space of merger results in a spin-up during minor merger.
Supporting our findings, recent observational studies have also found no clear correlation between post-merger features and spin properties \citep{Duc2015,Oh2016}.

To our surprise, the largest contribution to the spin evolution comes from ``Rest'' (blue curves). 
They may include extremely minor mergers (of ratios smaller than 1/50), secular evolution, fly-by encounters, harassment,
dynamical friction, and so on. 
For a sanity check we ran models of typical disk galaxies in ideal isolated environments and monitored their spin evolution, 
finding that their spin changes by less than 2\% over a Hubble time. 
Hence, numerical artifact is ruled out as a main contributor. 
Effects of gas accretion and star formation seem minimal too, 
as our simulation is on cluster environments where most of the constituent galaxies are cold-gas poor and quiescent.
Regarding the impact of extreme minor (larger than 1/500) mergers, we tested a case of massive galaxies (log $\rm{M}_{\ast}/{M}_{\odot} >$ 
10.5) and confirm that they do not make any substantial change in results.

Figure \ref{fig:f3}(C) may give us some hint as to the underlying cause. 
Model galaxies gradually spin down without mergers. 
Interestingly, some galaxies spin up occasionally. 
The slow rotating example in Figure \ref{fig:f3}(C) experiences spin up at $z \approx 1$, and it was caused by a fly-by interaction between this galaxy and the central galaxy of the cluster. 
Just like mergers, central-satellite fly-bys also have a high probability of reducing galaxy spin but can raise spin in the rare case of a corotating fly-by. 
It is difficult to predict the amount of spin evolution in cosmological environments analytically because such fly-by effects depend on many elements, such as halo merger history and the orbits of satellites. 
Another important issue is the difference between halo mergers and galaxy mergers. 
Many of the satellite galaxies in a cluster can be considered to be in the course of being merged. 
In practice, however, most of the satellite galaxies orbit around the cluster center for more than a Hubble time before finally merging. 
During such orbital motions of galaxies, various tidal effects occur and contribute to the spin evolution. 

Figure \ref{fig:f4} shows that the spin evolution is mass-dependent.
The overall total spin down ($\Delta\lambda \approx -0.25$) is similar between the two mass bins, but the details are different. 
The more massive sample experiences a larger spin down through mergers, and the combined effect of major and minor mergers is larger than that of ``Rest''. 
In the less massive galaxy sample, ``Rest'' dominates the spin evolution. It seems that different drivers operate the spin history in different mass regimes.
  
\section{Discussion} \label{sec:discussion}
Our models are for the moment exclusively on cluster galaxies, as our simulation was originally motivated by different questions. 
As a result, a direct comparison between the existing data and our models is not possible. 
In addition, the spatial resolution of our models is an issue. 
Our resolution is comparable to those of other state-of-the-art large-volume simulations but still too poor to adequately follow galaxy spin evolution. 
The resolution is barely good enough to resolve galactic disks and not sufficient to reproduce their vertical structure. 
For example, the scale height of the Milky Way stellar disk is roughly 300\,pc \citep{Cox2000} which is 2-3 times smaller than the maximum 
resolution of our models. 
The poor spatial resolution also makes it difficult to reproduce the radial disk structures such as bar and spiral arms. 
Such structures likely affect the disk rotation and the formation of spheroids as well. 
This means that our models are not currently capable of reproducing realistic disks. 
Galaxies are born as disks and (some) disk mergers lead to spheroids. 
Hence, without proper disks to begin with, proper spheroids may not be achieved either. 
Considering all these issues, our results on spin evolution should be taken only qualitatively. 

As a key to the history of galaxy formation and evolution, spin is considered more significant than ever before. 
Through the novel technique of IFU spectroscopy, 
we have access to the two-dimensional spectroscopic maps and thus the kinematic information of thousands and soon orders of 
magnitude more galaxies \citep{Croom2012, Sanchez2012, Bundy2014, Ma2014, Brodie2014, Bland-Hawthron2015}. 
Considering the highly non-linear nature of spin evolution, cosmological hydrodynamic simulations seem to be the best tool 
for its investigation. 

Using our cluster zoom-in simulations, we have gained a tantalizing first glimpse of the relative importance of the processes driving galaxy 
spin evolution. 
Galaxies do spin down through the cosmic history with and without mergers. Mergers are important yet not the dominant contributor. 
Many questions await answers. 
What is the ``Rest'' that seems the most significant operator in the spin evolution?
How can we pin down and quantify the contribution from each physical process while multiple processes act in complex combinations? 
Does the cluster mass or density in the environment matter? 
All these questions should and will be challenged in the near future through more extensive IFU observations and systematic (high resolution) simulation efforts.

\acknowledgments
We thank Rory Smith and Chandreyee Sengupta for reading the draft and useful comments.
SKY acknowledges support from the Korean National Research Foundation (NRF-2014R1A2A1A01003730).
The supercomputing time for numerical simulation was kindly provided by KISTI (KSC-2014-G2-003),
and large data transfer was 
supported by KREONET which is managed and operated by KISTI.
This study was performed under the umbrella of the joint collaboration between Yonsei University Observatory and the Korea Astronomy and Space Science Institute.

\bibliographystyle{aasjournal}

\begin{thebibliography}{}
\bibitem[{Aubert {et~al.}(2004)Aubert, Pichon, \& Colombi}]{Aubert2004}
Aubert, D., Pichon, C., \& Colombi, S. 2004, \mnras, 352, 376

\bibitem[{Barnes(1988)}]{Barnes1988}
Barnes, J. E. 1988, \apj, 331, 699

\bibitem[{Barnes(1992)}]{Barnes1992}
Barnes, J. E. 1992, \apj, 393, 484

\bibitem[{Behroozi {et~al.}(2013)Behroozi, Wechsler,\& Wu}]{Behroozi2013a}
{{Behroozi}, P.~S. and {Wechsler}, R.~H. and {Wu}, H.-Y. } 2013, \apj, 
762, 109

\bibitem[{Bertola \& Capaccioli(1975)}]{Bertola1975}
Bertola, F. \& Capaccioli, M. 1975, \apj, 200, 439

\bibitem[{Binney(1976)}]{Binney1976}
Binney, J. 1976, \mnras, 177, 19

\bibitem[{Bland-Hawthorn(2015)}]{Bland-Hawthron2015}
Bland-Hawthorn, J. 2015, in Proc. IAU Symp. 309, Galaxies in 3D across the
Universe, eds. B. L. Ziegler, F. Combes,  H. Dannerbauer, \& M. Verdugo (Cambridge: Cambridge Univ. Press), 21

\bibitem[{Bois {et~al.}(2011)Bois, Emsellem, Bournaud, Alatalo, Blitz, Bureau,
  Cappellari, Davies, Davis, de~Zeeuw, Duc, Khochfar, Krajnovi{\'{c}},
  Kuntschner, Lablanche, McDermid, Morganti, Naab, Oosterloo, Sarzi, Scott,
  Serra, Weijmans, \& Young}]{Bois2011}
Bois, M., Emsellem, E., Bournaud, F., {et~al.} 2011, \mnras, 416, 1654

\bibitem[{Brodie {et~al.}(2014)Brodie, Romanowsky, Strader, Forbes, Foster,
  Jennings, Pastorello, Pota, Usher, Blom, Kader, Roediger, Spitler, Villaume,
  Arnold, Kartha, \& Woodley}]{Brodie2014}
Brodie, J.~P., Romanowsky, A.~J., Strader, J., {et~al.} 2014, \apj, 796, 52

\bibitem[{Bundy {et~al.}(2014)Bundy, Bershady, Law, Yan, Drory, MacDonald,
  Wake, Cherinka, S{\'{a}}nchez-Gallego, Weijmans, Thomas, Tremonti, Masters,
  Coccato, Diamond-Stanic, Arag{\'{o}}n-Salamanca, Avila-Reese, Badenes,
  Falc{\'{o}}n-Barroso, Belfiore, Bizyaev, Blanc, Bland-Hawthorn, Blanton,
  Brownstein, Byler, Cappellari, Conroy, Dutton, Emsellem, Etherington,
  Frinchaboy, Fu, Gunn, Harding, Johnston, Kauffmann, Kinemuchi, Klaene,
  Knapen, Leauthaud, Li, Lin, Maiolino, Malanushenko, Malanushenko, Mao,
  Maraston, McDermid, Merrifield, Nichol, Oravetz, Pan, Parejko, Sanchez,
  Schlegel, Simmons, Steele, Steinmetz, Thanjavur, Thompson, Tinker, van~den
  Bosch, Westfall, Wilkinson, Wright, Xiao, \& Zhang}]{Bundy2014}
Bundy, K., Bershady, M.~A., Law, D.~R., {et~al.} 2014, \apj, 798, 7

\bibitem[{Cappellari(2002)}]{Cappellari2002}
Cappellari, M. 2002, \mnras, 333, 400

\bibitem[{Cox(2000)}]{Cox2000}
Cox, A. N. 2000, Allen's astrophysical quantities, (4th ed.; New York: AIP Press; Springer)

\bibitem[{Cox {et~al.}(2006)Cox, Dutta, Di Matteo, Hernquist, Hopkins, Robertson \& Springel}]{Cox2006}
Cox, T. J., Dutta, S. N., Di Matteo, T., {et~al.} 2006, \apj, 650, 791

\bibitem[{Croom {et~al.}(2012)Croom, Lawrence, Bland-Hawthorn, Bryant, Fogarty,
  Richards, Goodwin, Farrell, Miziarski, Heald, Jones, Lee, Colless, Brough,
  Hopkins, Bauer, Birchall, Ellis, Horton, Leon-Saval, Lewis, L??pez-S??nchez,
  Min, Trinh, \& Trowland}]{Croom2012}
Croom, S.~M., Lawrence, J.~S., Bland-Hawthorn, J., {et~al.} 2012, \mnras, 421, 872

\bibitem[{Dressler(1980)}]{Dressler1980}
Dressler, A. 1980, \apj, 236, 351

\bibitem[{Dubois {et~al.}(2012)Dubois, Devriendt, Slyz, \&
  Teyssier}]{Dubois2012}
Dubois, Y., Devriendt, J., Slyz, A., \& Teyssier, R. 2012, \mnras, 420, 2662
% 
\bibitem[{Dubois {et~al.}(2014)Dubois, Pichon, Welker, {Le Borgne}, Devriendt,
  Laigle, Codis, Pogosyan, Arnouts, Benabed, Bertin, Blaizot, Bouchet, Cardoso,
  Colombi, {De Lapparent}, Desjacques, Gavazzi, Kassin, Kimm, McCracken,
  Milliard, Peirani, Prunet, Rouberol, Silk, Slyz, Sousbie, Teyssier, Tresse,
  Treyer, Vibert, \& Volonteri}]{Dubois2014a}
Dubois, Y., Pichon, C., Welker, C., {et~al.} 2014, \mnras, 444, 1453

\bibitem[{Duc {et~al.}(2015)Duc, Cuillandre, Karabal, Cappellari, Alatalo,
  Blitz, Bournaud, Bureau, Crocker, Davies, Davis, De~zeeuw, Emsellem,
  Khochfar, Krajnovi??, Kuntschner, Mcdermid, Michel-Dansac, Morganti, Naab,
  Oosterloo, Paudel, Sarzi, Scott, Serra, Weijmans, \& Young}]{Duc2015}
Duc, P.~A., Cuillandre, J.~C., Karabal, E., {et~al.} 2015, \mnras, 446, 120

\bibitem[{Emsellem {et~al.}(1994)Emsellem, Monnet, \& Bacon}]{Emsellem1994}
Emsellem, E., Monnet, G., Bacon, R. 1994 \aap, 285, 723

\bibitem[{Emsellem {et~al.}(2007)Emsellem, Cappellari, Krajnovi{\'{c}}, {Van De
  Ven}, Bacon, Bureau, Davies, {De Zeeuw}, Falc{\'{o}}n-Barroso, Kuntschner,
  McDermid, Peletier, \& Sarzi}]{Emsellem2007}
Emsellem, E., Cappellari, M., Krajnovi{\'{c}}, D., {et~al.} 2007, \mnras, 379, 401

\bibitem[{Emsellem {et~al.}(2011)Emsellem, Cappellari, Krajnovi{\'{c}},
  Alatalo, Blitz, Bois, Bournaud, Bureau, Davies, Davis, de~Zeeuw, Khochfar,
  Kuntschner, Lablanche, Mcdermid, Morganti, Naab, Oosterloo, Sarzi, Scott,
  Serra, van~de Ven, Weijmans, \& Young}]{Emsellem2011}
Emsellem, E., Cappellari, M., Krajnovi{\'{c}}, D., {et~al.} 2011, \mnras, 414, 888


\bibitem[{Fall(1983)}]{Fall1983}
Fall, S. M. 1983, in Proc. IAU Symp. 100, Internal Kinematics and Dynamics of 
Galaxies, eds. E. Athanassoula 
(Springer), 391

\bibitem[{Fall \& Romanowsky(2013)}]{Fall2013}
Fall, S. M. \& Romanowsky, A. J. 2013, \apj, 769, 26

\bibitem[{Gerhard(1981)}]{Gerhard1981}
Gerhard O. E. 1981, \mnras, 197, 179

\bibitem[{Hernquist(1992)}]{Hernquist1992}
Hernquist L. 1992, \apj, 400, 460


\bibitem[{Hubble(1926)}]{Hubble1926}
Hubble, E. P. 1926, \apj, 64, 321

\bibitem[{Illingworth(1977)}]{Illingworth1977}
Illingworth, G. 1977, \apj, 218, 43


\bibitem[{Khandai {et~al.}(2015)Khandai, {Di Matteo}, Croft, Wilkins, Feng,
  Tucker, DeGraf, \& Liu}]{Khandai2015}
Khandai, N., {Di Matteo}, T., Croft, R., {et~al.} 2015, \mnras, 450, 1349

\bibitem[{Khochfar {et~al.}(2011)Khochfar, Emsellem, Serra, Bois, Alatalo,
  Bacon, Blitz, Bournaud, Bureau, Cappellari, Davies, Davis, de~Zeeuw, Duc,
  Krajnovi{\'{c}}, Kuntschner, Lablanche, McDermid, Morganti, Naab, Oosterloo,
  Sarzi, Scott, Weijmans, \& Young}]{Khochfar2011}
Khochfar, S., Emsellem, E., Serra, P., {et~al.} 2011, \mnras, 417, 845

\bibitem[{Komatsu {et~al.}(2011)Komatsu, Smith, Dunkley, Bennett, Gold,
  Hinshaw, Jarosik, Larson, Nolta, Page, Spergel, Halpern, Hill, Kogut, Limon,
  Meyer, Odegard, Tucker, Weiland, Wollack, \& Wright}]{Komatsu2011}
Komatsu, E., Smith, K.~M., Dunkley, J., {et~al.} 2011, \apjs, 192, 18

\bibitem[{Ma {et~al.}(2014)Ma, Greene, McConnell, Janish, Blakeslee, Thomas, \&
  Murphy}]{Ma2014}
Ma, C.-P., Greene, J.~E., McConnell, N., {et~al.} 2014, \apj, 795, 158

\bibitem[{Moody {et~al.}(2014)Moody, Romanowsky, Cox, Novak, \&
  Primack}]{Moody2014}
Moody, C.~E., Romanowsky, A.~J., Cox, T.~J., Novak, G.~S., \& Primack, J.~R.
  2014, \mnras, 444, 1475

\bibitem[{Naab {et~al.}(2006)Naab, Jessit, \& Burkert}]{Naab2006}
Naab, T., Jesseit, R., Burkert, A. 2006, \mnras, 372, 839


\bibitem[{Naab {et~al.}(2014)Naab, Oser, Emsellem, Cappellari, Krajnovi,
  McDermid, Alatalo, Bayet, Blitz, Bois, Bournaud, Bureau, Crocker, Davies,
  Davis, de~Zeeuw, Duc, Hirschmann, Johansson, Khochfar, Kuntschner, Morganti,
  Oosterloo, Sarzi, Scott, Serra, Ven, Weijmans, \& Young}]{Naab2014}
Naab, T., Oser, L., Emsellem, E., {et~al.} 2014, \mnras, 444, 3357

\bibitem[{{Oh} {et~al.}(2016){Oh}, {Yi}, {Cortese}, {van de Sande},
  {Mahajan}, {Jeong}, {Sheen}, {Allen}, {Bekki}, {Bland-Hawthorn}, {Bloom},
  {Brough}, {Bryant}, {Colless}, {Croom}, {Fogarty}, {Goodwin}, {Green},
  {Konstantopoulos}, {Lawrence}, {L{\'o}pez-S{\'a}nchez}, {Lorente}, {Medling},
  {Owers}, {Richards}, {Scott}, {Sharp}, \& {Sweet}}]{Oh2016}
{Oh}, S., {Yi}, S.~K., {Cortese}, L., {et~al.} 2016, \apj, 832, 690


\bibitem[{Peebles(1969)}]{Peebles1969}
Peebles, P. J.~E. 1969, \apj, 155, 393

\bibitem[{Sales {et~al.}(2012)Sales, Navarro, Theuns, Schaye, White, Frenk,
  Crain, \& {Dalla Vecchia}}]{Sales2012}
Sales, L.~V., Navarro, J.~F., Theuns, T., {et~al.} 2012, \mnras, 423, 1544

\bibitem[{S{\'{a}}nchez {et~al.}(2012)S{\'{a}}nchez, Kennicutt, {Gil de Paz},
  van~de Ven, V{\'{i}}lchez, Wisotzki, Walcher, Mast, Aguerri,
  Albiol-P{\'{e}}rez, Alonso-Herrero, Alves, Bakos, Bart{\'{a}}kov{\'{a}},
  Bland-Hawthorn, Boselli, Bomans, Castillo-Morales, Cortijo-Ferrero,
  de~Lorenzo-C{\'{a}}ceres, del Olmo, Dettmar, D{\'{i}}az, Ellis,
  Falc{\'{o}}n-Barroso, Flores, Gallazzi, Garc{\'{i}}a-Lorenzo, {Gonz{\'{a}}lez
  Delgado}, Gruel, Haines, Hao, Husemann, Igl{\'{e}}sias-P{\'{a}}ramo, Jahnke,
  Johnson, Jungwiert, Kalinova, Kehrig, Kupko, L{\'{o}}pez-S{\'{a}}nchez,
  Lyubenova, Marino, M{\'{a}}rmol-Queralt{\'{o}}, M{\'{a}}rquez, Masegosa,
  Meidt, Mendez-Abreu, Monreal-Ibero, Montijo, Mour{\~{a}}o, Palacios-Navarro,
  Papaderos, Pasquali, Peletier, P{\'{e}}rez, P{\'{e}}rez, Quirrenbach,
  Rela{\~{n}}o, Rosales-Ortega, Roth, Ruiz-Lara, S{\'{a}}nchez-Bl{\'{a}}zquez,
  Sengupta, Singh, Stanishev, Trager, Vazdekis, Viironen, Wild, Zibetti,
  Ziegler, V$\backslash$'ilchez, Wisotzki, Walcher, Mast, Aguerri,
  Albiol-P{\'{e}}rez, Alonso-Herrero, Alves, Bakos, Bart{\'{a}}kov{\'{a}},
  Bland-Hawthorn, Boselli, Bomans, Castillo-Morales, Cortijo-Ferrero,
  de~Lorenzo-C{\'{a}}ceres, del Olmo, Dettmar, D$\backslash$'iaz, Ellis,
  Falc{\'{o}}n-Barroso, Flores, Gallazzi, Garc$\backslash$'ia-Lorenzo,
  {Gonz{\'{a}}lez Delgado}, Gruel, Haines, Hao, Husemann,
  Igl{\'{e}}sias-P{\'{a}}ramo, Jahnke, Johnson, Jungwiert, Kalinova, Kehrig,
  Kupko, L{\'{o}}pez-S{\'{a}}nchez, Lyubenova, Marino,
  M{\'{a}}rmol-Queralt{\'{o}}, M{\'{a}}rquez, Masegosa, Meidt, Mendez-Abreu,
  Monreal-Ibero, Montijo, Mour{\~{a}}o, Palacios-Navarro, Papaderos, Pasquali,
  Peletier, P{\'{e}}rez, P{\'{e}}rez, Quirrenbach, Rela{\~{n}}o,
  Rosales-Ortega, Roth, Ruiz-Lara, S{\'{a}}nchez-Bl{\'{a}}zquez, Sengupta,
  Singh, Stanishev, Trager, Vazdekis, Viironen, Wild, Zibetti, \&
  Ziegler}]{Sanchez2012}
S{\'{a}}nchez, S.~F., Kennicutt, R.~C., {Gil de Paz}, A., {et~al.} 2012,
  \aap, 538, A8

\bibitem[{Schaye {et~al.}(2015)Schaye, Crain, Bower, Furlong, Schaller, Theuns,
  {Dalla Vecchia}, Frenk, Mccarthy, Helly, Jenkins, Rosas-Guevara, White, Baes,
  Booth, Camps, Navarro, Qu, Rahmati, Sawala, Thomas, \& Trayford}]{Schaye2015}
Schaye, J., Crain, R.~A., Bower, R.~G., {et~al.} 2015, \mnras, 446, 521

\bibitem[{Teyssier(2002)}]{Teyssier2002}
Teyssier, R. 2002, \aap, 385, 337

\bibitem[{{Toomre}(1977)}]{Toomre1977}
{Toomre}, A. 1977, in Evolution of Galaxies and Stellar Populations, ed.
  {B.~M.~Tinsley \& R.~B.~G.~Larson D.~Campbell} (New Haven, CT: Yale Univ. 
Observatory), 401

\bibitem[{{Tweed} {et~al.}(2009){Tweed}, Devriendt, Blaizot, Colombi, \& 
Slyz}]{Tweed2009}{Tweed}, D., Devriendt, J., Blaizot, J., Colombi, S., \& Slyz, 
A. 2009, \aap, 506, 647

\bibitem[{Vogelsberger {et~al.}(2014)Vogelsberger, Genel, Springel, Torrey,
  Sijacki, Xu, Snyder, Nelson, \& Hernquist}]{Vogelsberger2014a}
Vogelsberger, M., Genel, S., Springel, V., {et~al.} 2014, \mnras, 444, 1518
 
\end{thebibliography}

%% This command is needed to show the entire author+affilation list when
%% the collaboration and author truncation commands are used.  It has to
%% go at the end of the manuscript.
%\allauthors

%% Include this line if you are using the \added, \replaced, \deleted
%% commands to see a summary list of all changes at the end of the article.
\listofchanges

\end{document}